# Using a Binary Classification Model to Predict the Likelihood of Enrolment to the Undergraduate Program of a Philippine University

Joseph A. Esquivel[1], James A. Esquivel[2]

[1]*DIT Student, Graduate School, Angeles University Foundation*
*McArthur Highway, Angeles City, Philippines*

***Abstract** — With the recent implementation of the K to 12 Program, academic institutions, specifically, Colleges and Universities in the Philippines have been faced with difficulties in determining projected freshmen enrollees vis-à-vis decision-making factors for efficient resource management. Enrollment targets directly impacts success factors of Higher Education Institutions. This study covered an analysis of various characteristics of freshmen applicants affecting their admission status in a Philippine university. A predictive model was developed using Logistic Regression to evaluate the probability that an admitted student will pursue to enroll in the Institution or not. The dataset used was acquired from the University Admissions Office. The office designed an online application form to capture applicants' details. The online form was distributed to all student applicants, and most often, students, tend to provide incomplete information. Despite this fact, student characteristics, as well as geographic and demographic data based on the student's location are significant predictors of enrollment decision. The results of the study show that given limited information about prospective students, Higher Education Institutions can implement machine learning techniques to supplement management decisions and provide estimates of class sizes, in this way, it will allow the institution to optimize the allocation of resources and will have better control over net tuition revenue.*

**Keywords —** *Data Mining, Education Data Mining, Predictive Modeling, Binary Classification, Logistic Regression.*

## I. INTRODUCTION

Data mining is defined as the process of uncovering and analysis of large sum of data that targets to reveal hidden patterns and rules. Data mining supports in the extraction of knowledge from unprocessed data (Bra, et.al., 2012). In a study conducted by Sabnani, More, Kudale and Janrao (2018), they stated that data mining is the uncovering of hidden predictive information from database that includes the process of analyzing data from different perspectives and summarizing it into useful information to increase revenue. With the continuous growth of voluminous data, there is a good reason that smart data analysis will be more pervasive and will be a necessary component for technological advancement (Osisanwo et., al., 2017).

In the educational setting, Undaiva, Patel, Shah and Nikhil (2015), presented in their research that the main objective of Educational Data Mining (hereafter referred to as EDM) is to discover hidden patterns, association and relations to discover the hidden knowledge through different data mining techniques. The knowledge discovered by the techniques would enable the higher learning institutions to improve their educational processes which include making better decisions, having more advanced planning in directing students, predicting individual behaviors with higher accuracy, and enabling the institution to allocate resources and staff more effectively.

Higher Education Institutions exert much effort and substantial resources to influence, predict and recognize the decision-making choices of student applicants who have been offered admission (Basu, et. al., 2019). They are challenged by the uncertainty of human selection patterns, which greatly affects the target number of incoming students (Abelt, et. al., 2015).

The birth of K to 12 in the Philippines added up to the increasing pressure of the HEIs, college freshmen enrolment plunged to its lowest state in SY 2016-2017 that stretches to SY 2017-2018. Another growing concern is, graduates of the K to 12 Senior High School program (Grade 12), may or may not intend to enroll college since the SHS curriculum was designed to prepare students for one of the following: employment, entrepreneurship, Tech-Voc and Higher Education.

Predicting students' decision to pursue admission may help the University to save and properly manage resources and fairly distribute scholarship to underprivileged but deserving students.

## II. METHODOLOGY

This section illustrates the research design, procedures and planned data analysis used in the study.

### A. Freshmen Data Description

This section of the paper describes freshmen information which generally consists of information on their enrolment status, residency, gender, parent's job, type of Senior High School where they graduated from, School Choice.

Table I presents some of the demographic variables included in the study. The variables found in the dataset are considered as potential predictors of the dependent variable that is to enroll (Yes) or not enroll (No).

A total of nineteen (19) features were included in the Feature Selection process utilizing Weka Attribute Evaluator (supervised).

**TABLE I**
**STUDENT APPLICANT ADMISSION DATA**

| Feature | Description |
|---|---|
| OL Pursued | A binary variable indicating the whether the applicant is enrolled or not. The label is 1 if the applicant pursued enrolment and 0 otherwise. |
| Within_City | City or Municipality where the Applicant is residing |
| Within_Province | A binary variable which tells whether applicant lives within the province of Pampanga or not. |
| Religion_Binary | A binary variable which indicates the Religion of the applicant. Whether applicant is (0) Non-Catholic or (1) Catholic. |
| Type_of_School | A binary variable which indicates the type of SHS where the applicant graduated from. Whether the school is (0) Public or (1) Private. |
| With_Honors | A binary variable indicating whether the applicant received an honor or not. |
| School_choice | A binary variable which indicates whether the applicant's first school choice is the local institution. |

Table II displays the details of enrolees in terms of the number of applications, admitted and pursued enrolment for the school year 2018. A total number of 7,879 applications were received by the local university of which 4,486 were admitted and 3,414 pursued their enrolment. Table III presents the breakdown of enrolment based on the location of applicants' residences, whether within or outside the province where the University is located.

**TABLE III**
**STUDENT APPLICANT ADMISSION VS ENROLMENT**

| Applicants | Admit | Enroll | % |
|---|---|---|---|
| 7,879 | 4,486 | 3,414 | 76.10% |

**TABLE IIII**
**STUDENT APPLICANT RESIDENCE**

| Residence | Admit | Enroll | % |
|---|---|---|---|
| In-Province | 6,489 | 3,199 | 49.3% |
| Out-Province | 1,390 | 215 | 15.5% |

Table IV shows the gender breakdown of students, the percentage of those who applied and pursued enrolment were also indicated on the table.

**TABLE IVV**
**STUDENT APPLICANT GENDER**

| Gender | Admit | Enroll | % |
|---|---|---|---|
| Male | 3,635 | 1,870 | 51.4 |
| Female | 4,244 | 1,693 | 39.9 |

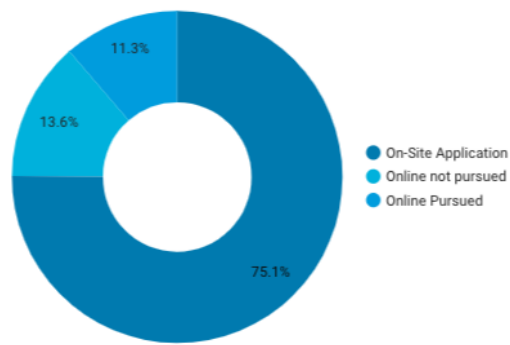

Fig. 1 shows the breakdown of students who applied online and those who opted to apply on-site. A total of 1,962 students applied on-line in which 892 pursued their enrolment and 1,070 did not. Most of the applicants preferred the traditional way of applying, on-site application process.

The graphs illustrated were generated using Google Data Studio. These figures enable the Institution's decision makers to quickly grasp the trend and may assist in making informed decisions for the school's main source of funds, students.

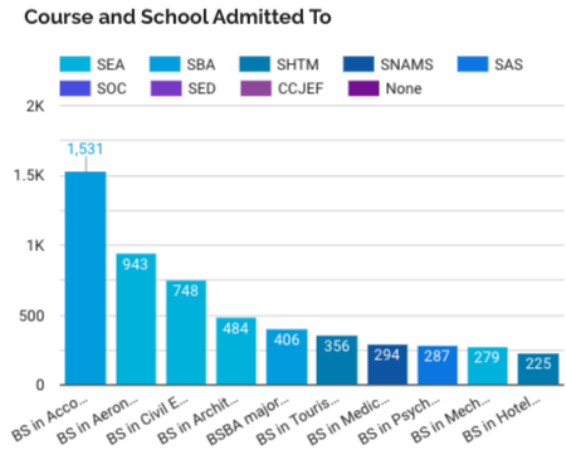

Fig. 2 indicates the breakdown of applicants courses and schools where they were admitted to. Accountancy registered the highest, followed by Aeronautics Engineering

*B. Research Design*

The regression model predicts a categorical dependent variable from a selected group of predictors (Fidell & Tabachnick, 2019). The predicted variable is how likely a student applicant will pursue enrollment in the institution. The study provided details about the data, discussed the preprocessing of the data, performed data exploration, and summarized the methodology. The researcher employed descriptive research design, since the study mined students' and University's characteristics. Distinguishing these attributes will help in predicting students' decision whether to pursue enrollment or not.

Prediction research design was also implemented to identify relationships of the features or variables with the dependent variable, to enroll or not to enroll. Since the study will predict the enrollment decision of a student, which is dichotomous in nature, binary classification algorithm, Logistic Regression, is implemented.

*C. Procedure*

Data mining process was applied to retrieve and transform the datasets which were sourced from the Admissions Office of the University. Initially, the dataset was cleaned using basic strategies such as implementing imputation and / or discard rows or columns containing missing values.

The data mining process involved the following steps: 1) enrollment data collection, 2) Data Cleaning and Wrangling 3) Feature Selection utilizing Correlation Attribute Evaluator and Ranker, 3) Splitting the records into Test and Train data, 4) train the classifier using k-fold cross validation, 5) execute model using test data set and evaluate the results. Dataset was prepared using MS Excel, cleaning involves removal of duplicates in data entry, dataset from different sources, Marketing Campaign and Actual Enrolment Records from the ITSS, were also combined in order to make meaningful insights, as seen in the tables and graphs. The combined dataset was processed using Weka in order to build the Logistic Regression model. The result of the model was evaluated based on its accuracy, sensitivity and specificity.

*D. Statistical Analysis of Data*

The study evaluated the prediction performance of the features, which is an essential part of the research undertaking. Classification method will allow to categorize the data into distinct number of classes. The classification algorithm used in the study is a Binary Classifier, also known as, Logistic Regression, with only two (2) possible outcomes, to enroll (Yes) or not enroll (No).

Confusion matrix table was also used to evaluate the accuracy of the prediction. Matrix was applied on the Logistic Regression results.

## III. RESULTS AND DISCUSSION

The purpose of this study was to create and apply a predictive model to predict college admission and enrollment using student characteristics. The study examined SY 2018 admission and actual enrollment datasets in a local university.

*A. Features Selection*

Features were selected using Classifier Subset Evaluator utilizing Logistic as the classifier in order to estimate the accuracy of the subsets and Best first as the search method utilizing bi-directional search direction.

The results, selected features, were then identified and applied to create the predictive model, see Figure 3. The dataset contains nineteen (19) features and among these features, twelve (12) features were selected, these selected attributes were then applied to the model in order to achieve the best accuracy percentage.

The following attributes with their corresponding rank, as shown in figure 3, were selected: OL Pursued, Within_City, Within_Province, Religion_Binary, College_Admitted_To_Binary, Guardian_Parent_Bnary, Total_Number_Siblings, Previous_School_Binary, With_Honors, School_Choice, Type_Of_School.

Table V presents the details of the performance metrics for evaluating the models.

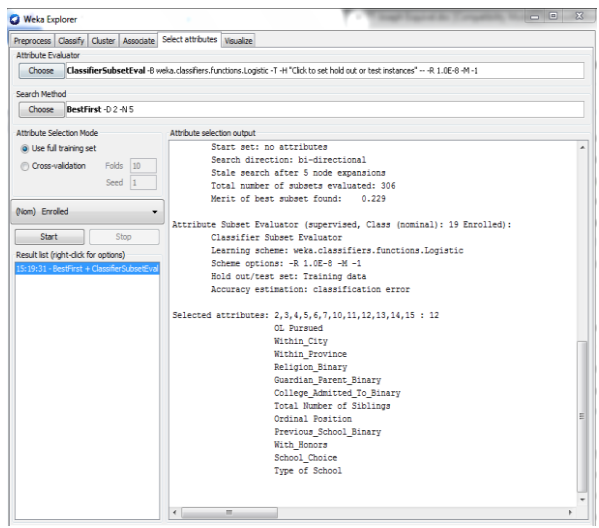

Fig. 3  Out of nineteen (19) features, twelve (12) were selected from the dataset using a bi-directional search method.

### B. Metrics for Evaluating the Model

The model was evaluated based on three performance measures: accuracy, sensitivity and specificity.  Table III shows the descriptions of the metrics.

**TABLE V**
**METRICS FOR EVALUATION**

| Regular | | |
|---|---|---|
| Metric | Description | Formula |
| Accuracy | Measures the ability of the model to correctly predict the class label of unseen data. | $\dfrac{TP + TN}{TP + TN + FP + FN}$ |
| Sensitivity | Measures the proportion of True Positives (or Yes's) that are correctly identified as such. | $\dfrac{TP}{TP + FN}$ |
| Specificity | Measures the proportion of True Negatives (or No's) that are correctly identified as such. | $\dfrac{TN}{TN + FP}$ |

### C. Logistic Regression Model

This section discusses the Logistic Regression results using Weka.  In building the model, as prescribed by Weka Attribute selector and to achieve a high accuracy rate, twelve (12) features were included. Also, all records in the dataset were used and ten (10) fold cross validation was applied in building the model.  Table VI below shows the Detailed Accuracy By Class.

**TABLE VI**
**METRICS FOR EVALUATION**

| Detailed Accuracy By Class | | | | |
|---|---|---|---|---|
| TP Rate | FP Rate | Precision | Recall | F-Measure |
| 0.843 | 0.327 | 0.757 | 0.843 | 0.798 |

True Positive (TP) Rate result, as shown above, is a fairly good indicator which means the model correctly predicted a label (predicted yes and it is a yes), False Positive (FP) Rate yield also have shown a good result, this is incorrectly predicting the label (predicted yes and it was actually no), the lower the value the better the interpretation.  Accuracy (F-Measure) shows an average of 79.8% accuracy rate with the selected features applied in the model.

### D. Confusion Matrix

This section shows an evaluation of the result using a confusion matrix.

**TABLE VII**
**CONFUSION MATRIX**

|  | Positive (1) | Negative (0) |
|---|---|---|
| Positive (1) | 3,637 | 679 |
| Negative (0) | 1,166 | 2,397 |

In order to test the performance of the LR model, the entire dataset was utilized as training set. Results from Weka showed that the model predicted 3,637 applicants were classified to enroll, on the other hand, 2,397 were predicted otherwise, not to enroll, in the Academic Institution, 679 applicants were incorrectly identified as not to enroll but will enroll and lastly, 1,166 applicants were incorrectly labeled.

### IV. CONCLUSIONS

The study illustrated the factors that influenced the enrolment decision of prospective students based on their characteristics. With the results generated, Machine Learning techniques may be used as a supplement to enrollment decisions and to measure the level of correlation between enrollment and such factors.  From nineteen (19) variables, Weka reduced the relevant features selected to twelve (12) which can yield a high accuracy performance.

The model was able to predict the enrollment of an applicant based on a given set of features with an accuracy rate of 80% using the selected attributes.

The accuracy rate percentage can be improved by adding other relevant features such as SHS GPA of applicants, NSAT Grades and Scholarships.

Future direction of the research is to integrate datasets from the Registrar, Marketing, Student Affairs and Testing Offices, comparison of multiple classifiction techniques other than LR such as Support Vector Machines (SVMs) and Neural Networks. Also, the implementation of Geocoding, that is to incorporate geographic locations on the model, like considering an applicant's location relative to that of the institution they are applying to. Lastly, the application of feature engineering, that is, designing surveys for prospective students to better understand their decisions for pursuing their enrolment to the University.

## ACKNOWLEDGMENT

The authors wish to acknowledge the assistance extended by the participating institution especially the offices of the Admissions, University Registrar and the ITSS.